\documentclass[12pt,twoside]{article}
\usepackage{fleqn,espcrc1}
\usepackage{graphicx,epsfig}

\newcommand{\AmS}{{\protect\the\textfont2
  A\kern-.1667em\lower.5ex\hbox{M}\kern-.125emS}}
\title{Thermalization and Flow in 158 $A$GeV Pb+Pb Collisions}
\author{H. Schlagheck\address{Institut f\"ur Kernphysik, 
        Westf\"alische Wilhelms Universit\"at, \\ 
        Wilhelm Klemm Str. 9, 48149 M\"unster, Germany}
        for the WA98 collaboration}
\begin{document}
\maketitle
\begin{abstract}
  The WA98 experiment at the CERN SPS measures Pb+Pb Collisions 
at 158 $A$GeV. 
  The scaling properties of the charged particle multiplicity
at midrapidity with the number of participants are studied using the
SPMD and the MIRAC detector. Neutral pion spectra obtained from 
the LEDA detector are compared to hydrodynamical parametrizations.
The collective flow in the target fragmentation region has been studied 
using the Plastic Ball detector. The results exhibit a strong dependence 
on centrality and rapidity.
\end{abstract}

\section{Introduction}
Ultra-relativistic heavy-ion collisions produce dense matter which 
is expected to form at sufficiently high energy densities a deconfined 
phase of quarks and gluons, the Quark Gluon Plasma. A necessary condition
for such a phase transition is local equilibrium which might be 
achievable through rescattering of produced particles. Hints for 
thermalization can most easily identified by studying the observables
as function of centrality. The high $p_T$ 
pion production is expected to be dominated by hard parton scattering but 
has recently been shown to be also explainable by a thermal model with 
hydrodynamic expansion. The comparison of the WA98 $\pi^0$ data 
with hydrodynamic models provides constraints on the partition of 
excitation energy in terms of temperature and an average flow velocity.
Such a finite thermalized system without any external pressure will
necessarily expand and the thermal pressure will generate 
collective motion which will be reflected in the momentum spectra
of the final hadrons. Thus a part of the thermal excitation energy will 
be converted into collective motion of the hadrons.

\section{Scaling of Global Variables}
\begin{figure}[t]
   \centerline{\psfig{figure=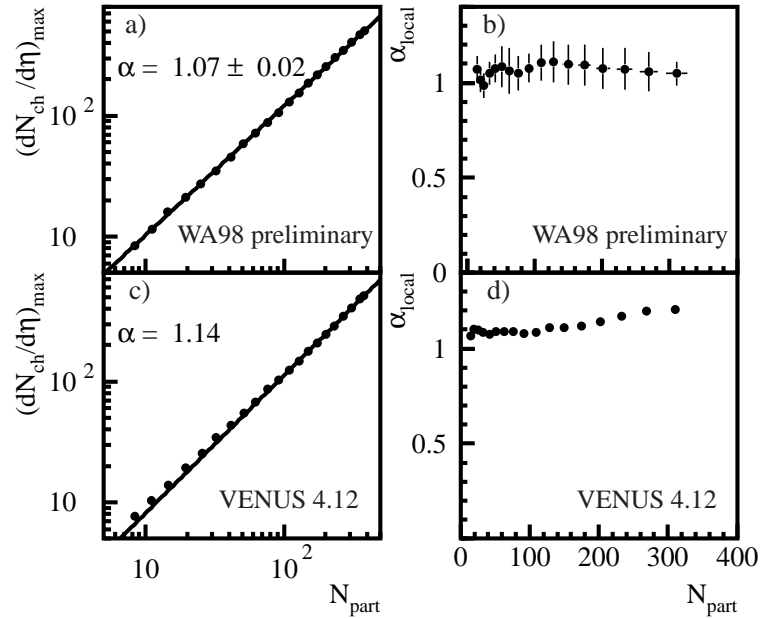,width=10cm}}
    \vspace{-1cm}
   \caption{\protect\label{scaling} Pseudorapidity density of $N_{ch}$ (a) at
            midrapidity as a function of the number of participants.
            Figure b) shows the local scaling exponents which have 
            been obtained from a fit of 5 neighbouring data points. 
            Figures c)~-~d) show similar results obtained from 
            from VENUS 4.12 simulations. }
    \vspace{-0.5cm}
\end{figure}  
There seem to be qualitative changes in the behaviour of heavy-ion 
reactions once a certain system size is attained. Strangeness production
is enhanced in S+S reactions compared to p+p, but seems to saturate for 
even larger nuclei\cite{QM97-SOLL-97-A}. Recent results from the WA98
experiment\cite{WA80-AGGA-98-A} show significant change of the shape of 
the $\pi^0$ $p_T$ spectrum in peripheral Pb+Pb collisions compared to 
p+p data. However, from semi-central Pb+Pb reactions with about 50 
participating nucleons up to the most central reactions the shape 
remains unchanged.
In the present analysis the centrality has been selected with the
transverse energy $E_T$ measured in the calorimeter MIRAC.
 Fig.~\ref{scaling} shows the
scaling behaviour of the charged particle multiplicity $dN_{ch}/d\eta$
with the number of participants $N_{part}$. It can be seen 
that $dN_{ch}/d\eta$  follows a power law with the 
number of participants. The extracted exponent from the data is 
$\alpha = 1.07$. On the bottom part of fig.~\ref{scaling} the same 
analysis performed with VENUS 4.12 \cite{SIM-WERN-93-A} is displayed. 
While the simulation result also obeys roughly a power law scaling, 
the agreement is not as good, and the scaling exponent appears to be 
significantly larger than that obtained from the experimental data.

\section{Neutral Pion Spectra}
The neutral meson spectra are mainly influenced by thermal and 
chemical freeze-out in the final state.
In the analysis of central reactions of Pb+Pb at 158 $A$GeV it is 
seen that both predictions of perturbative QCD 
\cite{WA80-AGGA-98-A,SIM-WANG-98-D} and hydrodynamical parameterizations 
\cite{WA80-AGGA-98-F} can describe the measured 
neutral pion spectra very well. 
It is particularly astonishing to observe that on the one hand 
the pQCD calculation gives a good description also at relatively low momenta 
while on the other hand the hydrodynamical parameterization would yield 
a sizable contribution even at very high momenta.
\begin{figure}[t]
        \centerline{\psfig{figure=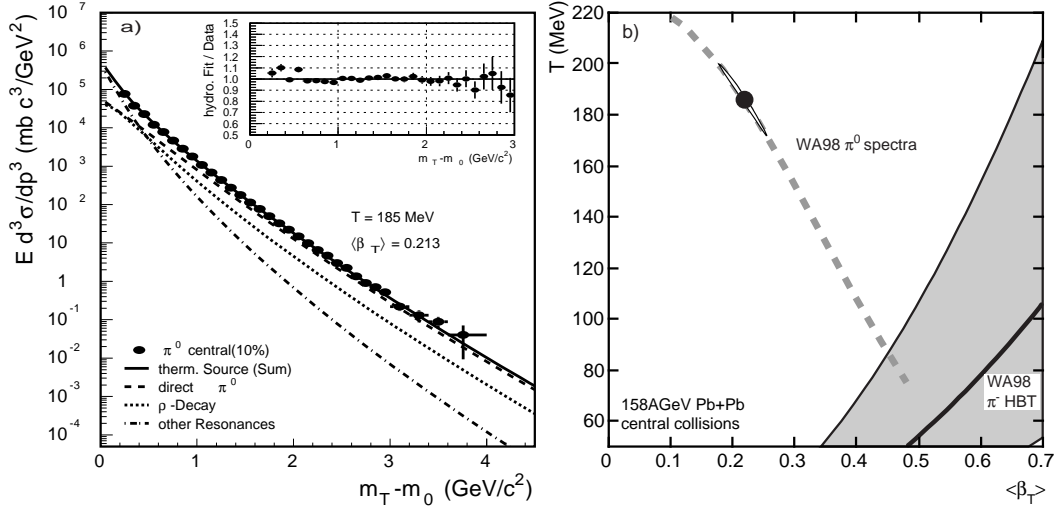,width=14cm}}
    \vspace*{-1cm}
        \caption{a) Transverse mass spectra of neutral pions in central 
        collisions of 158 $A$GeV Pb+Pb.
        b) Freeze-out parameters $T$ and $\langle \beta_{T} \rangle$ for fits 
        to the neutral pion spectra. The best fit is shown as a solid 
        circle and the corresponding 2-$\sigma$-contour is displayed as a 
        thin black line. The grey dashed band indicates the possible parameter 
        sets if one of the two is constrained to a fixed value. 
        The broad grey band shows the 
        1-$\sigma$ allowed region from pion interferometry.
        \protect\label{hydro}}
    \vspace*{-.5cm}
\end{figure}

Fig.~\ref{hydro}a shows a comparison of the neutral pion 
spectra to a fit of a hydrodynamical parameterization including transverse 
flow and resonance decays \cite{SIM-WIED-97-B}. Using the default 
Gaussian profile the best fit is obtained with $T = 185 \, 
\mathrm{MeV}$ and $\langle \beta_{T} \rangle = 0.213$. 
Fig.~\ref{hydro}b shows the best fit parameters as a filled 
circle -- the corresponding $2 \sigma$ contour is also shown. The figure 
also contains the $1 \sigma$ allowed region from the $m_{T}$ 
dependence of the transverse radii extracted by negative pion 
interferometry with the WA98 negative tracking arm \cite{QM99-VORO-99-A}. 
The interferometry constraints are very similar 
to those given in \cite{NA49-APPE-97-B} -- they favour relatively large 
transverse flow velocities. Such large velocities are only compatible with 
the neutral pion spectra, if one assumes a very different spatial 
profile. However, this would result in rather low temperatures, thus 
these parameters are very sensitive to the used profile\cite{QM99-PEIT-99-A}.

\section{Collective Flow}
If the initial state of the evolution is azimuthally asymmetric, as in 
semi-central heavy-ion collisions, this property will be reflected in 
the azimuthal asymmetry of the final state particle distributions. 
The strength of the collective flow will yield
information on the nuclear equation of state during the expansion.
Collective flow development follows the time evolution of pressure 
gradients in the hot, dense matter. Thus, collective flow can serve
as a probe to provide information on the initial state and to which 
extent the reaction zone might me thermalized. In particular,
the formation of a Quark Gluon Plasma during the early stages of the 
collision is expected to result in reduced pressure gradients due to
a softer nuclear equation of state which results in a reduced collective
motion\cite{SIM-RISC-96-B}.

Since the Plastic Ball Detector in the WA98 experiment is azimuthally 
symmetric it is ideal to perform the analysis of azimuthal anisotropies.
\begin{figure}[t]
\vspace*{-5mm}
\begin{minipage}[t]{7cm}
\psfig{figure=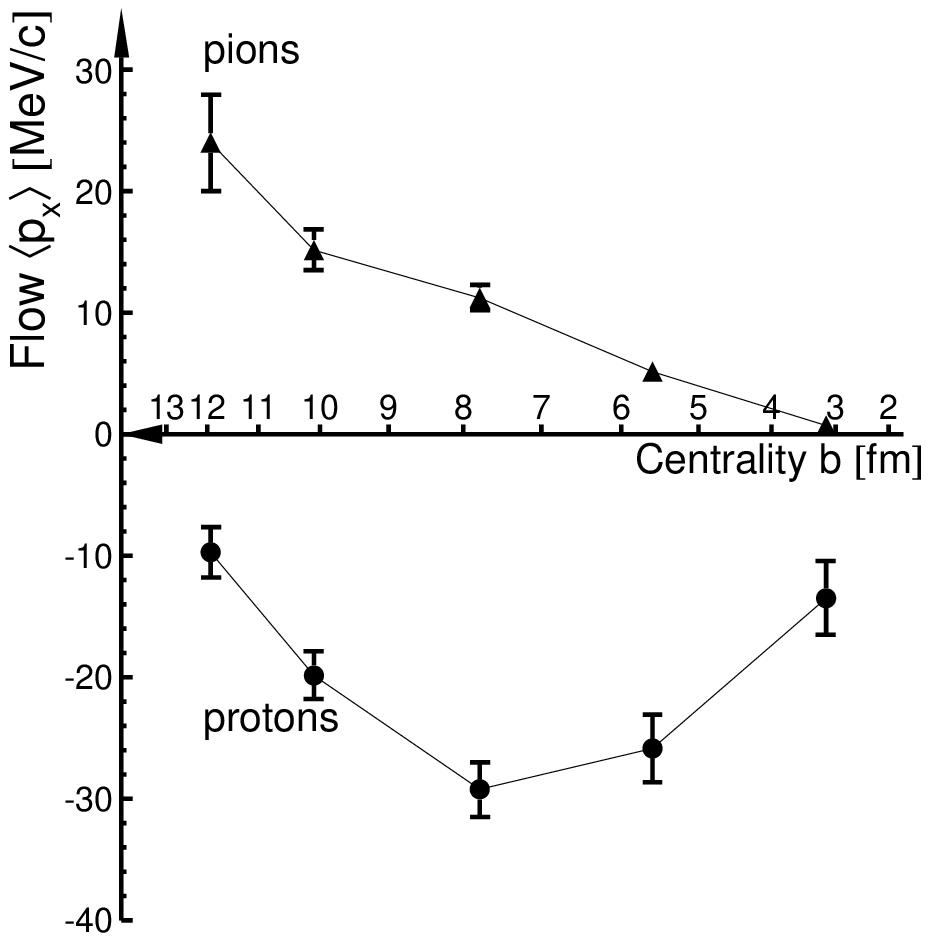,width=77mm}
\vspace*{-15mm}
\caption{\label{flowcent} The average transverse momentum as function 
of centrality in terms of the impact parameter $b$.}
\end{minipage}
\hspace{5mm}
\begin{minipage}[t]{7cm}
\psfig{figure=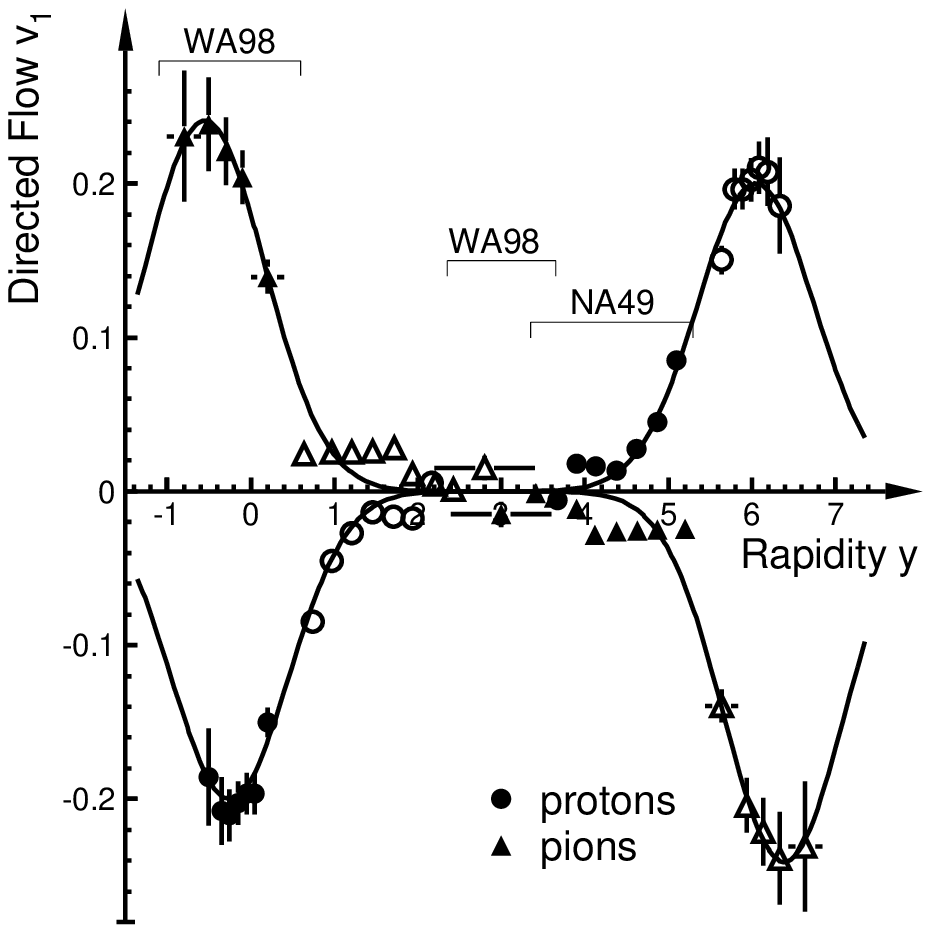,width=77mm}
\vspace*{-15mm}
\caption{\label{flowy} The flow parameter $v_1$ as function of rapidity $y$ 
in semi central collisions.}
\end{minipage}
\vspace*{-5mm}
\end{figure} 
Fig.~\ref{flowcent} shows the centrality dependence of the directed
flow in terms of the average transverse momentum $\langle p_x \rangle$.
For protons the maximum directed flow is observed in reactions with 
intermediate centrality. The corresponding impact parameter of 
$b \approx 8 \mathrm{fm}$\cite{QM99-SCHL-99-A} results twice as large 
as observed for AGS energies\cite{E877-BARR-94-A}. Since the observed
$\langle p_{x} \rangle$ of pions is positive it indicates that 
the pions are preferentially emitted away from the target spectators,
this is called anti-flow\cite{SIM-JAHN-94-A}.
The rapidity dependence of directed flow is given in fig.~\ref{flowy}.
 In addition, pion data measured with the tracking arm
in the WA98 experiment\cite{PANIC-NISH-99-A} at midrapidity and data
near midrapidity measured by the NA49 collaboration\cite{NA49-APPE-97-A}
are shown.  The maximum flow is observed in the fragmentation regions, 
while it rapidly decreases near midrapidity. The data follow a 
Gaussian distribution.

  Hence for a complete description of the rapidity distribution of the 
collective flow the formerly used slope at midrapidity
 (e.g. $d\langle p_x \rangle / dy|_{y=0}$)
is not sufficient. It is more reasonable to use the three parameters of 
the Gaussian distribution to describe the data. The peak position reflects 
the beam momentum, the peak height gives the strength of the flow and the 
width of the distribution provides information on how much the participants 
and the spectators are involved in the collectivity.

\end{document}